# Reporting an Experience on Design and Implementation of e-Health Systems on Azure Cloud

Shilin Lu, Peter Strazdins

Research School of Computer Science
Australian National University
Canberra ACT 0200 AUSTRALIA
{s.lu, peter.strazdins}@anu.edu.au

Rajiv Ranjan

Information Engineering Laboratory
CSIRO ICT Centre
Canberra ACT 0200 AUSTRALIA
raj.ranjan@csiro.au

*Abstract*—**Electronic Health (e-Health) technology has brought the world with significant transformation from traditional paper-based medical practice to Information and Communication Technologies (ICT)-based systems for automatic management (storage, processing, and archiving) of information. Traditionally e-Health systems have been designed to operate within stovepipes on dedicated networks, physical computers, and locally managed software platforms that make it susceptible to many serious limitations including: 1) lack of on-demand scalability during critical situations; 2) high administrative overheads and costs; and 3) in-efficient resource utilization and energy consumption due to lack of automation. In this paper, we present an approach to migrate the ICT systems in the e-Health sector from traditional in-house Client/Server (C/S) architecture to the virtualised cloud computing environment. To this end, we developed two cloud-based e-Health applications (Medical Practice Management System and Telemedicine Practice System) for demonstrating how cloud services can be leveraged for developing and deploying such applications. The Windows Azure cloud computing platform is selected as an example public cloud platform for our study. We conducted several performance evaluation experiments to understand the Quality Service (QoS) tradeoffs of our applications under variable workload on Azure.**

## I. INTRODUCTION

Providing efficient healthcare services is becoming a challenging problem for governments across the world. On one hand, the increasing health-awareness among people has led to soaring demands for the health care services. On the other hand, the governments have limited fund and limited personnel to dedicate to this sector. Recent studies have shown that the health care sector consumes a large of proportion of GDP in many countries. Over the past decades, ICT has been widely adopted within the health care sector, which has significantly improved the work efficiency [23]. This has led to the highly developed e-Health care service sector worldwide. e-Health is a new term where ICT systems are deployed for better management and coordination of information. Marconi [7] quotes: 'e-health is an application of the Internet and other related technology in the healthcare sector to improve the access, efficiency, effectiveness and quality of clinical and business processes utilized by health care organizations, practitioners, patients, and consumers in an effort to improve the health status of patients'.

However, the substantial application of ICT systems to healthcare has led to number of serious concerns [21]. Since large number of computing, storage and networking equipments are widely deployed for e-Health applications in the hospitals, the equipment consume huge electrical power or energy, which is becoming an issue. According to Chamara [8], a traditional PC Dell 2350 1.8GHz Pentium 4 (only the host unit) consumes 6Watts in its sleep state and 60 to 85Watts when fully powered on. Another study by Lawrence Berkeley National Laboratory suggested that 60% of all desktop PCs in commercial buildings remain fully powered-on, 36% were turned off and 4% were asleep during nights and weekends [9] with existing power management utilities of the computing servers or networking equipment almost always disabled [8]. Further, the research has confirmed that even at a very low load, such as 10% CPU utilization, the power consumed is over 50% of the peak power [3]. Similarly, non-consolidated ICT systems also lead to increased cooling costs.

Recent technological advances in e-Health services, such as Medical Body Area Networks (MBAN) are challenging the existing in-house ICT infrastructures. According to the market intelligence company ABI research [2], over the next five years, close to five million disposable wireless MBAN sensors will be shipped. MBANs enable a continuous monitoring of patient's condition by sensing and transmitting measurements such as heart rate, electrocardiogram (ECG), body temperature, respiratory rate, chest sounds, and blood pressure etc. MBANs will allow: (i) real-time and historical monitoring of patient's health; (ii) infection control; (iii) patient identification and tracking; and (iv) geo-fencing and vertical alarming. However, to manage and analyze such massive MBAN data from millions of patients in real-time, healthcare providers will need access to an intelligent and highly scalable ICT infrastructure.

Hence, it is clear that there is an immediate need to leverage efficiency and dynamically scalable ICT infrastructure for deploying current and next-generation eHealth applications [32]. We propose to achieve this by leveraging cloud computing systems [1][26][31]. Cloud computing assembles large networks of virtualised ICT services such as hardware resources (such as CPU, storage, and network), software resources (such as databases, application servers, and web servers) and applications. In industry these services are referred to as Infrastructure as a Ser-vice (IaaS), Platform as a Service (PaaS), and Software as a Service (SaaS). Cloud computing services are hosted in large data centres, often referred to as data farms, operated by companies such as Amazon, Apple, Google and Microsoft.

Today, cloud computing presents an immense opportunity for the health care sector. First, it can significantly reduce the initial capital investments in the IT infrastructure in hospitals due to the pay-as-you-go pricing models [24]. Second, it can improve the utilization of IT resources and improve the quality of health care service delivery among the healthcare sector [21]. In addition, sharing and managing large amounts of distributed medical information including EHR and X-Ray images across the e-Health system through cloud environment is the current trend [18] [26]. The cloud storage services provides the a good and scalable solution for such massive data management challenges [25].

We note that optimising energy efficiency [30] of cloud computing data centre has also emerged as one of the important research in the past few years. Discussion on how to optimize the energy efficiency of data centres is beyond the scope of this paper. We assume that cloud data centre provider implements number of software and hardware-based approaches to perform energy optimizations. *Our argument is based on the fact that cloud computing has better energy efficiency than traditional C/S application hosting approaches, as it does better consolidation of application workload via dynamic system scaling and de-scaling (server shutdown, migration, and the like)*

In this paper we present two e-Health applications, which are programmed as SaaS applications using Azure cloud services at PaaS and IaaS layers. We developed two practical applications (Cloud-based Medical Practice Management System [27] and Cloud-based Telemedicine Practice System [28]) to demonstrate how cloud computing can be applied for e-Health for overcoming those limitations on traditional ICT architectures and improving the scalability and energy efficiency of healthcare applications. The Windows Azure cloud computing platform is carefully selected as an example of public cloud platform for hosting our applications. On the programming level, ASP.NET MVC programming model, SQL Azure Database and C# programming language are leveraged for developing our applications.

The main contributions of the paper are: (1) we illuminate a concept of migrating the e-Health applications from traditional C/S architecture to cloud computing environment for improved energy-efficiency and salability; (2) we present designs and implementations of two e-Health applications and their deployments in a public (Microsoft Azure) cloud computing environment; and (3) we conduct extensive experiments for evaluating the Quality of Service (QoS) features of our application on Windows Azure.

The rest of the paper is organized as follows. In section II, the literature review will be given to show the state-of-the-art e-Health and cloud computing research. In section III, we will discuss details in the e-Heath applications and cloud-based ICT architecture deployed in the hospitals. In section IV, we will demonstrate the outcome of experiments. We end the paper in section V with concluding remarks.

## II. LITERATURE REVIEW

Computers have been widely used by health practitioners since 1990's. Nowadays, most doctors, nurses and other health practitioners are using personal computers to process the patients' records, prescriptions, and appointments. Some typical computing approaches are widely adopted by the health care sector such as EHRs, e-Prescription, e-Pharmacy and telemedicine. The term Electronic Health Record (EHR) refers to the digital records saved in database to store the patients' personal medical information. The records can contain various types of information, such as patients' personal profile, physiological data, medical history, prescriptions from the medical providers, physiological test results, or even some multimedia data such as digital X-ray films. The EHR data can be used for further verification of patient's condition by the doctors, or provided to the insurance for claim verification. Compared with paper-based medical records used in the past, the EHR system has many advantages, including easy to search and store.

Many countries have initiated high profile HER programs. For example, in Australia, the government has appointed the National E-Health Transition Authority Limited (NEHTA) to research the EHR system since 2004. Now, this system is expected to be launched by 2013 [12]. In China, the EHR system has been developed for many years it will be deployed by next year. Moreover, in contemporary hospitals and clinics, electronic prescription (e-prescription) has been prevailing as another popular application in e-Health practice, so is the electronic pharmacy (e-Pharmacy). Telemedicine is another contemporary approach for connecting the patient and doctor at distance using high definition video conferencing technologies. It is usually deployed with videoconference device, audio device, scanner and respective data compression algorithm to transfer the data between two points. The benefit of this approach is obvious. According to O'Reilly [4], in Alberta, Canada, an evaluation of telepsychiatry services has been shown to improve the satisfaction for the both the patients and the psychiatrists, as it saved the travel time.

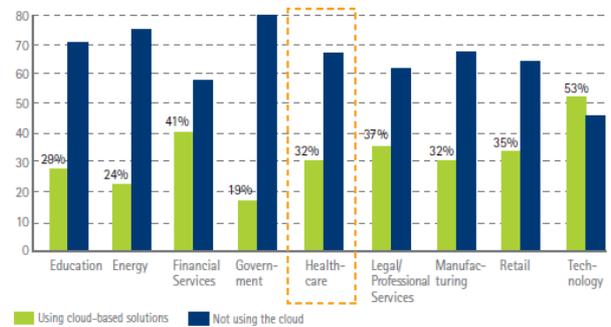

**Table 1:** Cloud computing adoption across different sectors [25]

In addition, the cloud computing technique is good for health data exchange, data mining for health science research. In the traditional way, in order to discover new drugs and new medical treatments, scientists need to analyze vast medical data over years where millions of dollars are invested. With cloud computing, investments and time consumption are significantly reduced [24]. Further, the cloud can be used for body sensor network, for example, the patients' physiological data at remote can be accessed and verified by the doctors over the cloud [22] [25] in real time. Obviously, in recent years, more and more sectors including healthcare are adopting cloud computing to replace their own computing infrastructure. A survey indicates that almost three-quarters of health industry respondents are planning or already using cloud-based services [25]. Table 1 shows 32% sectors are acting to move to cloud platform from the traditional computing facilities.

## III. APPLICATION ARCHITECTURE

In order to demonstrate the advantages of hosting the health applications in cloud computing environments, we developed two e-Health applications and deployed them on the Windows Azure cloud platform. Though there are several vendors, who provide cloud computing services, including Amazon, IBM, Google and Microsoft. Compared with other cloud platform, Windows Azure has a few unique advantages. The first advantage is that the Visual Studio .Net Development Platform has close integration Windows Azure cloud platform leading to seamless application programming and deployment experience. Therefore, one can develop and debug cloud-based applications within Visual Studio. Microsoft also provides a tool called 'Publish to Windows Azure' for developer to deploy the application to Azure. Another advantage of Windows Azure is that it directly integrates the 'SQL Azure' as the database system into the cloud platform. Developers can move their SQL server database from existing in-house database systems to Azure cloud with minimal programming. The third advantage is that Windows Azure provides very high performance backup mechanism for database, web service, virtual machine and virtual network.

Our first application is called 'Cloud-based Medical Practice Management System' and the second is called 'Cloud-based Telemedicine Practice System'. Figure 1 shows the Windows Azure-based system architecture for deploying the proposed e-Health applications. From the figure, we can see that there are three layers in our cloud-based e-HaaS application architecture, IaaS, PaaS and SaaS. At the IaaS layer, Windows Azure provides the infrastructure services such as virtual machines, storage unit and high speed network.

Our application components are deployed on the virtual machines to provide web services as well as to store the huge health data on the Azure's Blob storage, which is a service for storing large amounts of unstructured data that can be accessed via HTTP or HTTPS. A single blob can be very big in size (Gigabytes), and a single storage account can contain up to 100TB of blobs which can be used for distributed access [19]. Windows Azure provides two different kinds of blobs; first one is block blob and second one is page blob. Hence, Azure can support different data formats based on type and mix of e-Health applications. Our applications leverage blob storage to save the patients' image files, X-ray data, EHR documents, video and audio data. Assuring security [29] and privacy [33] of e-Health data on public cloud storage services is beyond the scope of this paper. However, in future we intend to integrate our e-Health applications with TrustStore system [29] developed by CSIRO. TrustStore is a service-oriented solution for provisioning hybrid (including both private and public) data centre resources with strong guarantees on data security and privacy.

On the user side or client side based on the IaaS layer, all the typical desktop computers in the hospital can be replaced with the thin cloud terminals. In this case, all the healthcare staff can directly access the cloud-hosted applications via remote desktop Utility.

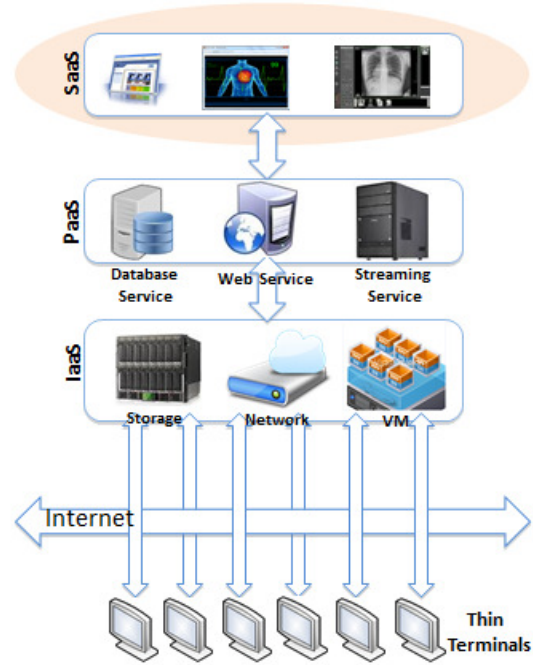

**Fig. 1.** Windows Azure-based e-Health Application Architecture

At the PaaS layer, our applications rely on the web services to provide online services, the SQL Azure server to provide SQL queries service and the media server to provide live streaming service for the videoconference communication. Our two e-Health applications operate at SaaS layer for providing the software functionalities for the users or healthcare staff (doctors, nurses, pharmacists, etc.). Our application can be scale based on the workload demands (e.g., number of users, data size, etc.). Discussion on how our applications can be automatically scaled dynamically based on workload demands is beyond scope of this paper.

### A. Cloud-based Medical Practice Management System

The first e-Health application is called 'Cloud-based Medical Practice Management System', which is shown in Figure 2. The application is integrated with the typical health management systems which can be used for the hospitals, clinics and other medical organizations. In this system, most useful medical relevant business processes and data are efficiently managed including EHR, e-Prescription, Personal Health Archives, X-Ray data, e-Pharmacy Management, e-Appointment, Billing, Accounting and Finance management systems.

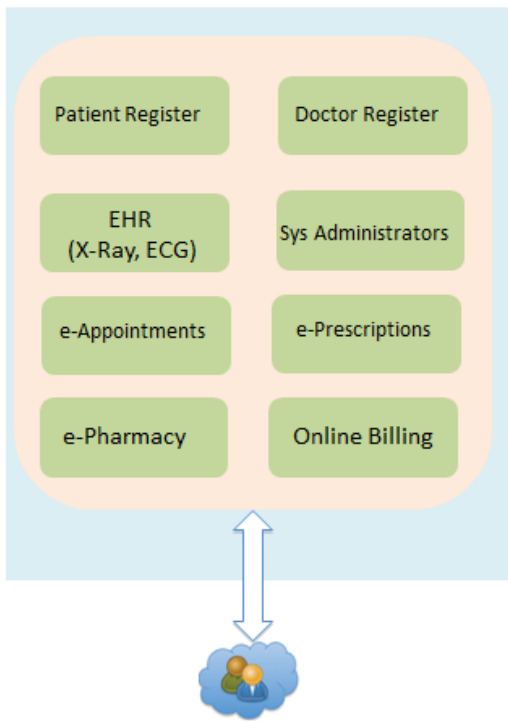

**Fig. 2.** Cloud-based Medical Practice Management System

In this application, we developed the software components using C# language on the Visual Studio 2012 development platform. We adopt the MVC programming model and SQL Server 2008 as the database. Figure 3 shows a UML (Unified Modeling Language) flow diagram of the Cloud-based Medical Practice Management System. In this flow diagram, the main business processes are described as four models: patient module, doctor module, system administrator module and pharmacy module.

Since the application is designed using MVC model, there several core classes defined in the Controller component. These core 9 classes are shown in the UML diagram in Figure 4. Each class implements a service. In each class, few methods are defined as well based on their relative functionalities.

In the pharmacy service, the 3 core sub-module is defined as DrugInventory, DrugPurchase and DrupSales. For example, in the DrugInventory class, 6 methods are created as Figure 5: (1) DrugInventoryCreate; (2) DrugInventoryDetails; (3) DrugInventoryIndex; (4) DrugStockCreate; (5) DrugStockDetails; (6) DrugStockIndex.

In this application, the database is MS SQL and deployed on the SQL Azure database system of Windows Azure platform. After create a SQL database instance on Azure, users can run the SQL Server scripts which is developed on local machine. The database will be replicated on one SQL Azure instance. Application can connect to the database instance on Azure SQL server through the ADO.NET or ODBC connection strings. In this application, several tables are designed. Figure 6 shows the UML database diagram of Cloud-based Medical Practice Management System. There are 4 main database tables are designed to support the most functionalities in the e-Health system including 'Patients', 'eAppointments', 'Billings' and 'EHR'.

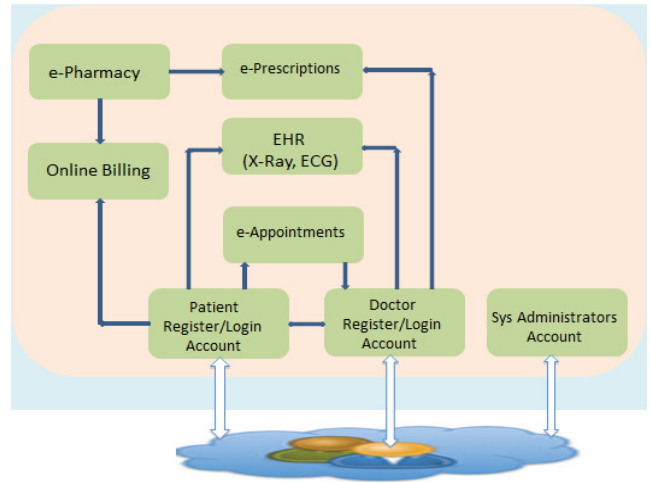

**Fig. 3.** Cloud-based Medical Practice Management System

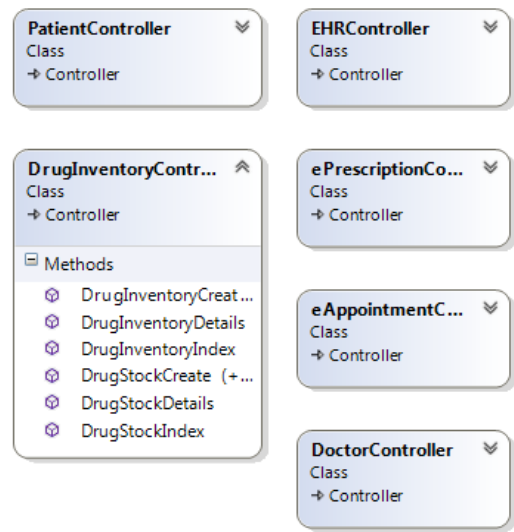

**Fig. 4.** Classes defined in Controller componnet

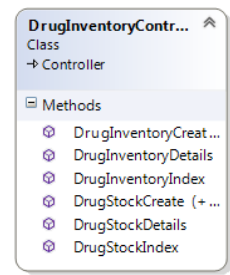

**Fig. 5.** Methods defined in DrugInventory Class

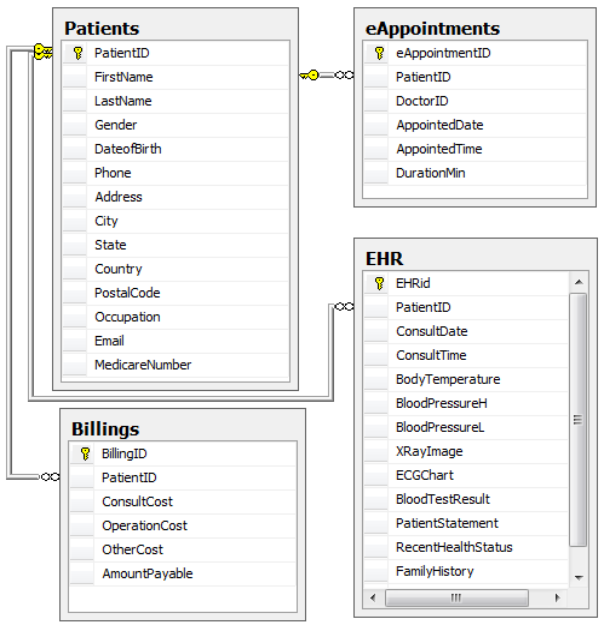

**Fig. 6.** SQL Azure Database Diagram

## B. Cloud-based Telemedicine Practice System

Another e-Health application applied on the cloud architecture is called 'Cloud-based Telemedicine Practice System'. It is integrated with multi-functions such as e-Appointment, e-Consulting, Telemedicine and e-Prescription. Based on the Azure cloud platform, the patients can see the doctors by remote through internet and consult any health problems. The doctor can check the health records, X-Ray graphics etc. for the patient and even check the patient's live physiologic data by real-time with some body sensor network.

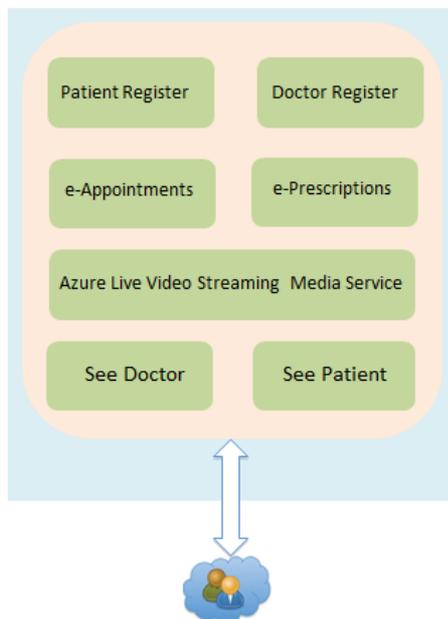

**Fig. 7.** Cloud-based Telemedicine Practice System

Telemedicine application is a hybrid system. It combines diverse communication techniques and hardware. In this experiment, we adopt web camera as the user front device and Microsoft Expression Encoder 4 as the video compression technique, Windows Azure virtual machine as the live video stream server. In the Azure cloud platform, the live streaming media server is deployed for video/audio processing. Figure 7 shows the diagram of main modules provided in the Cloud-based Telemedicine Practice System. Window Azure offers the scalable video streaming service through the web role media service running in the virtual machine. In the up streaming process of Figure 8, the Expression Encoder 4 will be used to collect the raw data from the camera on the client (either a patient or a doctor) and encoder the raw video streaming to Silverlight format; meanwhile, the streaming will be pushed to a Windows Azure web role which is a virtual machine. The web-role is an IIS media service which can receive the video streaming through a push point, for instance, domain.com/push.isml/mainfest. In the down streaming process, a Windows Azure Web Role loads the video streaming by balance and publishes it on the pull point, for example, domain.com/pull.isml/manifest. This process is executed by the Media Server on the Windows Azure Web Role. This publish point enables the video player to pull video stream from the Media Server. On the client end, the Silverlight player should be installed with web browser, either Internet Explorer or Firefox. The Silverlight formatted video stream will be decoded and show by the web browser.

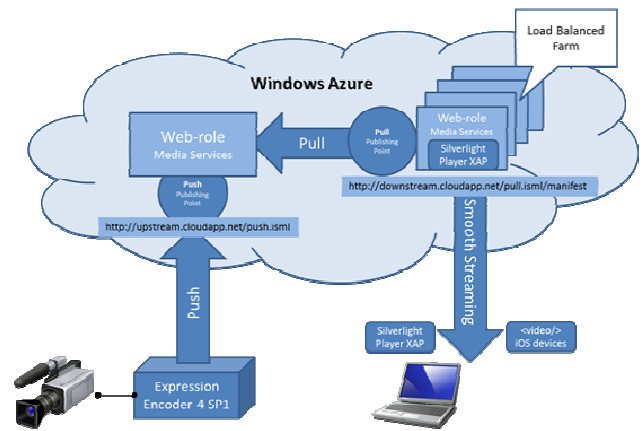

**Fig. 8.** Windows Azure Live Smooth Streaming [20]

Figure 9 shows the work flow of the telemedicine system. In the system, a patient can register and then login to to search the appropriate doctor at remote for him/her and then make appointment with the remote doctor on the available date and time. Then at the appointed date and time, the patient login the system to meet with the appointed doctor. The patient is able to see the doctor on the system interface by video and audio and then talk to the doctor with his/her statement. In addition, the patient can present his/her other physiological data to the doctor, such as his/her recent blood test report, X-ray image, and EGG chart or other multimedia data or text descriptions. At the end of consultation, the doctor can give the conclusion and solution or prescription. The prescription will be presented to the patient on his/her system. The patient will follow the advice. Then the consultation ends.

## IV. PRELIMINARY EXPERIMENTAL RESULTS

In our experiments, we deploy the aforementioned e-Health applications on Windows Azure cloud [35]. The following services on Windows Azure are leveraged for this deployment: Cloud Web Services, Azure SQL Databases, Virtual Machine, IIS and Live Smooth Streaming Media Services. We assumed unsaturated server availability for these experiments, so that enough capacity could always be allocated to a virtualised web service or SQL database for any service request. At user-end, we simulate large number of users in the hospital, who access the services of two e-Health applications using different workload generation tools. First, we use JMeter [36] for simulating 20 groups of concurrent users to evaluate the scalability of our applications hosted on Windows Azure cloud. In particular, we quantify the Average Response Time of HTTP Request (ARTHR) of the following components related to the Cloud-based Medical Practice Management System: (i) web service and (ii) SQL database.

Figure 10 shows the measurement for the ARTHR under different-sized groups of concurrent users. As expected, from the initial results it is clear that the ARTHR degades as the number of users accessing the service increase. We observed that the ARTHR stayed below 200ms for a user population below 93. However, as the concurrent users workload was increased to 93 or above, the ARTHR grows beyond 200ms. For example, as we increased the concurrent user poupluation to 1024, the ARTHR rises significantly and finally it soars to 1205ms. The red line on the graph shows the elevated HTTP response latency. A red horizontal line at 200ms shows the threshold of the HTTP response latency of web service on Azure cloud. That means our e-Health applications with its current Azure configuration (single instance) provides the best QoS interms of ARTHR for the web service and database query when we keep the concurrent user below than 93. We believe that our applications can be scaled for much larger population of users via implementation of an autonomic application provisioning technique, which will adapt to the increase or decrease in workload by dynamically scaling the number of instances of the application components. We intend to investigate this aspect in our future work.

In the second experiment we evaluate the scalability and QoS of the live video streaming service hosted on the virtual machine (an IaaS layer service) of Windows Azure. In the Cloud-based Telemedicine Practice System, we deploy the Live Smooth Streaming Media Services on the virtual machine to enable the patient and doctor to implement an easy to use online medical consultation via the video conferencing system. In this experiment, we use Smooth Stream Performance Testing Tool [37] that generates simulated workload for the streaming media services. This tool simulates the real user, who try to connect to the media stream available from the streaming service. Here we analyse the average video/audio chunk retrieval time for assessing the media streaming QoS of the virtualised media service.

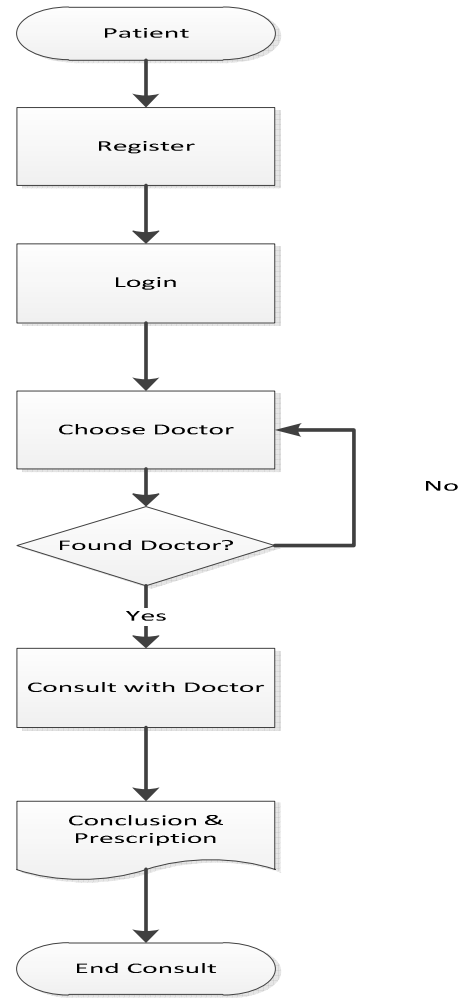

**Fig. 9.** Work flowchart of Telemedicine Practice System

In this experiment, we selected a medium size virtual machine with 2 cores, 3.5GB memory and located at the datacenter in West US for hosting the media service. In our system, the video conferencing involves 2 users, i.e., one patient and other the healthcare provider. Our early results show that the performance of audio stream remains contant over different chunks or files as shown in Figure 11. This is understandble as the audio files are not generally network communication heavy if compressed using an efficient encoding technique. However, the video stream shows uncertain QoS in terms of response time for different chunks or file fragments, as shown in Figure 12. We believe that uncertain video QoS is due to the variability of network bandwidth between the two users. In future work, we will work on developing network QoS profiling technique for dynamically learning the congestion and bandwidth between users and cloud-hosted applications. Such a technique will help us in dynamic migration of video content across data centres for improving user's QoS.

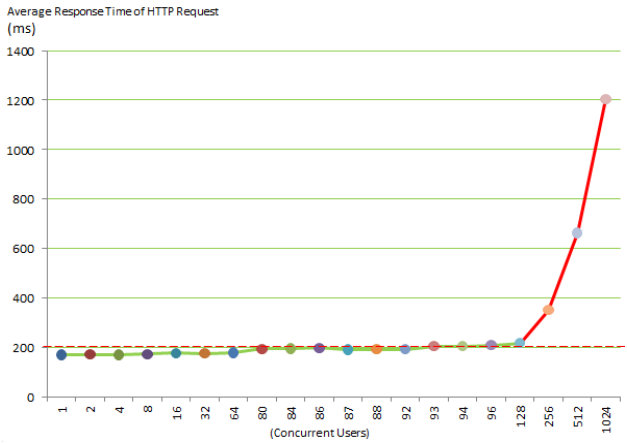

Average Response Time of HTTP Request (ms)

**Fig. 10.** Average Response Time of HTTP Request to concurrent users

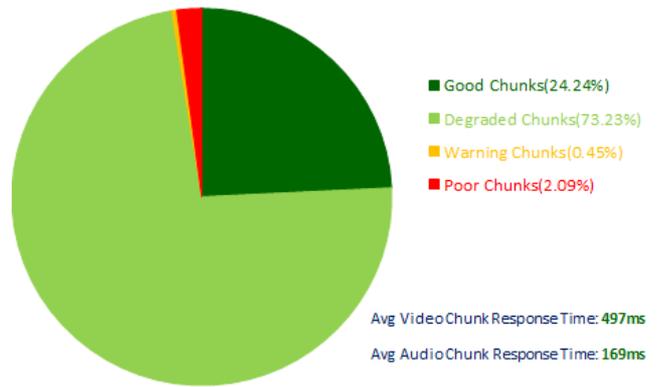

Good Chunks(24.24%)
Degraded Chunks(73.23%)
Warning Chunks(0.45%)
Poor Chunks(2.09%)

Avg Video Chunk Response Time: **497ms**
Avg Audio Chunk Response Time: **169ms**

**Fig. 13.** Media Stream Services Performance with 4 Cameras

| Chunk Type | Bitrate | Fragment Index | Response Time |
|---|---|---|---|
| audio | 64kbps | 1040719229 | 164ms |
| audio | 64kbps | 1022143173 | 164ms |
| audio | 64kbps | 1003102810 | 165ms |
| audio | 64kbps | 980231156 | 165ms |
| audio | 64kbps | 964093287 | 164ms |
| audio | 64kbps | 940408934 | 163ms |
| audio | 64kbps | 921832925 | 163ms |
| audio | 64kbps | 900354467 | 163ms |
| audio | 64kbps | 882823354 | 163ms |
| audio | 64kbps | 860067799 | 166ms |
| audio | 64kbps | 841491836 | 164ms |
| audio | 64kbps | 820477778 | 165ms |
| audio | 64kbps | 802017914 | 164ms |

**Fig. 11.** Average Audio Chunk Response Time

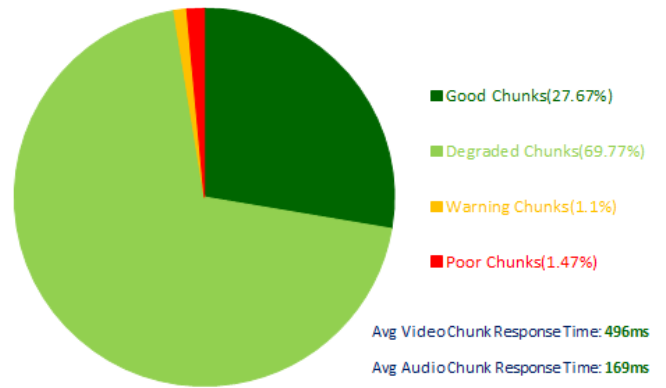

Good Chunks(27.67%)
Degraded Chunks(69.77%)
Warning Chunks(1.1%)
Poor Chunks(1.47%)

Avg Video Chunk Response Time: **496ms**
Avg Audio Chunk Response Time: **169ms**

**Fig. 14.** Media Stream Services Performance with 2 Cameras

Figure 13 and 14 show the media stream service's performance with 4 cameras and 2 cameras respectively. We can see the average video chunk response time is 497ms at the scenario of 4 cameras compared with 496ms at the scenario of 2 cameras. While the audio performace is the same 169ms during the two scenarios.

| Chunk Type | Bitrate | Fragment Index | Response Time |
|---|---|---|---|
| video | 400kbps | 2191619705 | 488ms |
| video | 400kbps | 2166821321 | 489ms |
| video | 400kbps | 2142002474 | 486ms |
| video | 400kbps | 2117395074 | 489ms |
| video | 400kbps | 2093106349 | 658ms |
| video | 400kbps | 2068467064 | 488ms |
| video | 400kbps | 2044179014 | 488ms |
| video | 400kbps | 2019691348 | 490ms |
| video | 400kbps | 1995403057 | 490ms |
| video | 400kbps | 1971084462 | 487ms |
| video | 400kbps | 1946445346 | 164ms |
| video | 400kbps | 1921676889 | 975ms |
| video | 400kbps | 1897038189 | 487ms |

**Fig. 12.** Average Video Chunk Response Time

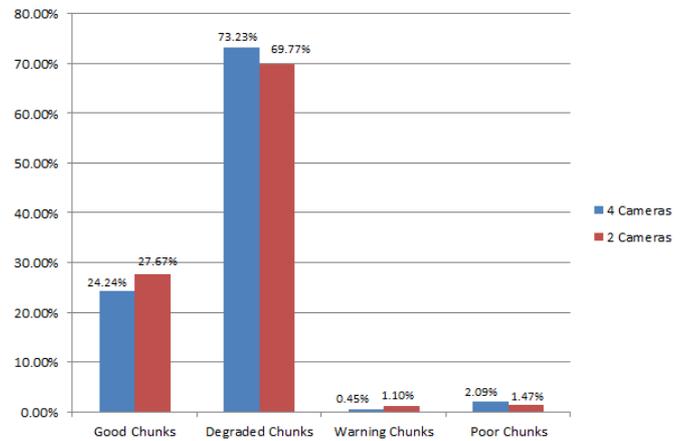

**Fig. 15.** Comparison of the Media Stream Services Performance

In Figure 15, we compare the overall performance of media stream services for two scenarios. We can see there is no significant differences in the two situations due to different video content delivery workload generated by cameras. As expected for 4 cameras setting the performance of the media stream service degraded to increasing processing and network transfer overload. Overall, we conclude that it is feasible to engineer e-Health applications using public cloud services. However, to guarantee QoS of applications one needs to develop intelligent cloud service and network provisioning technique. Further, one also needs to develop data security and privacy preserving techniques for protecting confidential e-Health data.

## V.  CONCLUSION

In this paper, we presented our experience on designing and implementing two e-Health application systems by leveraging Azure cloud platform. We started by analyzing challenges (lack of scalability, energy in-efficiency and the like) healthcare sector faces when using traditional C/S architecture for delivering e-Health services. Our proposed approach addresses these challenges by leveraging cloud computing services. We have implemented the prototype of our e-Health application systems and successfully evaluated its QoS performance on Azure cloud under variable workload settings. As part of our ongoing work, we are working on an intelligent elastic framework for autonomic provisioning of e-Health applications in private or public cloud environments. This framework will allow knowledge-driven optimised resource provisioning where it adapts to uncertain data streams/volumes, the number of users and varying resource and workload unpredictability.

## ACKNOWLEDGMENT

Thanks to Zheng Li and Miranda Zhang. They provided some valuable advices and enthusiastic assistances for the experiments.